\documentclass[12pt]{article}
\usepackage{sbc-template}
\usepackage{graphicx,url}

\usepackage[utf8]{inputenc}

\title{Review of LoRaWAN Applications}

\author{Lucas R. de Oliveira\inst{1}, Poliana de Moraes\inst{1}, Lauro P. S. Neto\inst{1}, and \\Arlindo F. da Conceição \inst{1}}

\address{Universidade Federal de São Paulo
  (UNIFESP), 
  São J. dos Campos, Brazil
\email{\{restivo.lucas,poliana.moraes,lauro.paulo,arlindo.conceicao\}@unifesp.br}
}

\begin{document} 

\maketitle

\begin{abstract}
  This paper presents a systematic review of LoRaWAN applications. We analyzed 71 cases of application, with focus on deploy and challenges faced. The review summarizes the characteristics of the network protocol and shows applications in the context of: smart cities, smart grids, smart farms, health, location, industry, and military.  Finally, this article analyzes some security issues.
\end{abstract}
  
\section{Introduction}

Low Power Wide Area Network (LPWAN) is a low-power, low-cost, wide area coverage and low data transfer wireless communication, designed to Internet of Things (IoT)~\cite{b1}. Examples of LPWAN technologies are LoRa and LoRaWAN~\cite{LoRa}.

LoRa is a physical layer modulation technique, based on Chirp Spread Spectrum, and patented by the French company Semtech. It can be used in the unlicensed radio frequency spectrum for data transmission, including ISM band. LoRaWAN is a network protocol convergent to the LPWAN and defines the system architecture.

This systematic review seeks to highlight the potential of using LoRa and LoRaWAN in real IoT applications. Previous studies~\cite{b88} pointed out some characteristics of the technology, but the practical scenarios and considerations about deploy have not yet been explored. 
To explore these aspects, this paper, in Section~\ref{lora}, presents the general characteristics of the technology.
Section~\ref{met} summarizes the methodology used in this systematic review. In Section~\ref{des}, the selected articles are organized in categories according to their area of application. 
Section~\ref{sec}, reports some security issues and concerns. 
Finally, Section~\ref{concl} concludes the study.

\section{LoRa and LoRaWAN protocols} 
\label{lora}

\begin{figure} [ht!]
\centering
\includegraphics[width=.4\textwidth]{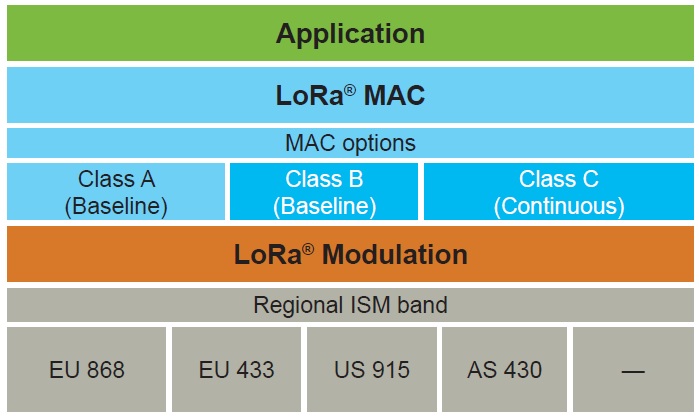}
\caption{Layers of the LoRaWAN protocol.}
\label{fig:pilha}
\end{figure}

The architecture and specification of LoRa and LoRaWAN can be found in LoRaWAN Alliance portal. The Figure \ref{fig:pilha} presents the layered organization, at the top level are the applications. It is also worth mentioning that, at bottom level, are the regional specifications.
These definitions are based on local regulations in each country that determine the best allocation of the radio transmittable space, since it is essential to have the fewest collisions and especially that such transmissions do not interfere in operations considered critical, such as communications to public security and patrimony of such governments. On account of these definitions, some countries also determine that a maximum duty cycle value (1\% in most countries) is respected, which is crucial for good data flow and fair usage of airtime.

Regarding the LoRa transmission settings, the Spreading Factor (SF) is a firmware-configured integer variable that varies from 7 to 12. The variable SF is used to change the modulation of the information to be transmitted according to the needs of each application. By using the same SF for the same channel may cause packet collisions;  but different SFs for the same channel do not present problems due to the immunity to interference, thats is an important characteristic of the technology. Thus, to choose the right SF parameter have a direct consequence in two questions: air time and distance of reach~\cite{b10}.

In addition, the SF parameter has influence in the maximum payload. In LoRaWAN, due to the duty cycle restrictions, by using SF7 the maximum payload size is 242 bytes. It is 51 bytes with SF12~\cite{b11}.

\section{Systematic review methodology} \label{met}

The methodology adopted in this review~\cite{b53} is organized as follows. Initially, in Phase 1, IEEE, ACM, Springer, ScienceDirect digital libraries were used for search (keywords: ``LoRaWAN'', ``LoRa'', ``LPWAN'', ``IoT'' and ``sensor'' with the AND connector). In Phase 2, the title and abstract were analyzed for suitability. In Phase 3, primary reading was performed and,  in Phase 4, we applied some exclusion criteria. Articles that met all the requirements were carefully analyzed in Phase 5.

Regarding the exclusion criteria, all papers published before 2015 (when LoRa version 1.0 was released) and not written in English were excluded. During the last phase, Phase 5, all articles were into Application, Theoretical and Experimental groups, as shown in Table \ref{tab1}. The group of \textit{Application} articles contemplated all the cases in which LoRa helped in the solution of some real situation. The \textit{Theoretical} was formed by studies  without implementation in the field. The \textit{Experimental} cases were composed of articles that underwent tests with LoRa, but also did not envisage practical application.

\begin{table}[htbp]
\caption{Number of articles per database}
\begin{center}
\begin{tabular}{|p{2.5cm}|p{2cm}|p{1.5cm}|p{2cm}|p{2cm}|p{1.5cm}|}
\hline
\textbf{Type of work} & \textbf{\textit{IEEE Xplore}}& \textbf{\textit{ACM}}& \textbf{\textit{Science Direct}}& \textbf{\textit{Springer Link}} & \textbf{Total}\\
\hline
\textbf{Application}& 59& 4& 4& 4& \textbf{71}\\
\hline
\textbf{Theoretical}& 51& 4& 9& 7 & \textbf{71}\\
\hline
\textbf{Experimental}& 22& 3& 2& 2& \textbf{29}\\
\hline
\textbf{Total}& 132& 11& 15& 13& \textbf{171}\\
\hline
\end{tabular}
\label{tab1}
\end{center}
\end{table}

This paper is focused on articles from the first group, the real applications. We were interested in investigate deploy, equipment used, and perceived problems in practice.


\begin{table}[htbp]
\caption{Country of origin}
\begin{center}
\footnotesize
\begin{tabular}{|p{3cm}|p{10cm}|}
\hline
\hline
\textbf{Local}& \textbf{References}\\
\hline
\hline
Argentina & \cite{b24, b83}\\
Australia&\cite{b27}\\
Australia/Russia &\cite{b23}\\
Belgium&\cite{b35, b85}\\ 
Brazil&\cite{b33,b34,b37,b72}\\
Bulgaria&\cite{b48,b63,b80}\\
China&\cite{b64}\\
China/England&\cite{b22}\\ 
Czech Republic&\cite{b82}\\ 
Denmark&\cite{b44,b66}\\ 
England&\cite{b32}\\
Finland&\cite{b1,b15}\\
France&\cite{b26,b36,b45}\\ 
France/Ireland&\cite{b81}\\
Germany&\cite{b74}\\
Germany/Tunisia&\cite{b42}\\ 
Greece&\cite{b40}\\
India&\cite{b25,b68,b52,b60,b67}\\
Italy&\cite{b71,b56,b11,b38}\\
Italy/Sweden&\cite{b47}\\ 
Italy/Turkey&\cite{b41}\\
Japan &\cite{b78}\\
Kazakhstan/Russia &\cite{b49}\\
Lithuania&\cite{b87}\\
Macedonia&\cite{b65}\\
Malaysia&\cite{b20,b21,b75,b76,b79}\\
Morocco&\cite{b30}\\
Netherlands&\cite{b58}\\
Pakistan&\cite{b70}\\
Romania&\cite{b10}\\
Scotland&\cite{b28}\\
Serbia/Slovenia&\cite{b43}\\
Singapore/Myanmar&\cite{b73}\\
South Korea&\cite{b55, b62}\\
Taiwan&\cite{b29,b46,b57,b69,b86}\\
Thailand&\cite{b13,b50}\\
Tunisia&\cite{b84}\\
Turkey&\cite{b59}\\
Uruguay&\cite{b39}\\
USA&\cite{b31,b77}\\
USA/Finland&\cite{b61}\\
USA/Portugal&\cite{b12}\\
\hline 
\hline 
\end{tabular}
\label{tab2}
\end{center}
\end{table}


\subsection{Research questions}

The research questions used during Phase 5 were:
\begin{enumerate}
\item What are the main features of LoRa technology?
\item What applications are LoRa and LoRaWAN being used for??
\item What are the main challenges that LoRa faces today?
\end{enumerate}

\section{ Results} \label{des}
The list of 71 analyzed  articles is, ordered by country, in Table \ref{tab2}. Malaysia, India, and the state of Taiwan have 5 articles each. In Brazil, four articles were identified. Some articles were developed in partnership between universities of different countries.

\subsection{Question 1: what are the main features of LoRa technology?}\label{AA}

Several papers did analogies among LoRa, LoRaWAN and OSI network model, where LoRa integrates the physical layer and LoRaWAN the network layer~\cite{b12}. Besides these analogies, an article clarified the classes of LoRa~\cite{b13}. According to the study, \textbf{Class A} is intended for simple sensors and for better battery utilization.
\textbf{Class B} is intended for actuators. And \textbf{Class C} is for powered devices that need low latency but with reduced battery efficiency, they are bi-directional and always active.
Most of the current LoRa applications is based on Class A. 

The throughput can be between 0.3 kbps and 50 kbps, considering the different spreading factor configurations for 915 MHz and the number of 64 channels for uplink and downlink~\cite{b14}. The battery for a stand-alone system can remain active for up to 10 years~\cite{b57}.

About the range distance, the expected range is 5 km for urban areas and up to 14 km for rural areas, or areas without significant obstacles \cite{b14}, and 30 km on the water surface \cite{b15}. In Bucharest, Romania~\cite{b10}, a hardware was prototyped to validate the maximum transmission distances reached using LoRa. They reached, using SF7, a range of 4.3 km in urban area and 9.7 km in rural areas, with the gateway antenna positioned in high places. Besides, it was reported that an Arduino would not be able to operate as a real gateway because it only possessed 2 kB of SRAM.





\subsection{Question 2: what applications are LoRa and LoRaWAN being used for?}

The practical use cases are detailed below, organized by area of application:

\begin{itemize}

\item \textbf{Applications for Smart Cities}. In South Korea, a network that reaches 99\% of the population is being developed using LoRaWAN, while Amsterdam, Netherlands, was covered entirely with 10 gateways \cite{b32, b58}. London and surrounding areas, England, are already adequately covered by LoRa~\cite{b22}. The total project cost of 11,681 nodes and 47 gateways exceeded one million pounds, for a financial return in up to 7 years. Around 90\% of the investment was explained by the costs of installing, maintaining and leasing local infrastructure.

There are several applications in public transport and traffic monitoring. In India \cite{b25}, some buses are equipped with transceivers and bus stops with receivers (nRF24L01 transceiver integrated to PIC18F and ESP8266). In Nagoya, Japan~\cite{b78}, it was possible to track the location of buses using such transceivers at bus stops with a LoRa AL-050 next to an Arduino Uno and GPS U-blox NEO-7N to capture bus positioning in time real. In Argentina \cite{b24}, Cordoba and Buenos Aires are cities that already have LoRa networks implemented with nodes made of BeagleBone Black, GPS, and LoRa module. In Selangor, Malaysia, with the aim of improving the traffic monitoring in denser urban areas in Malaysia, where there are intense vehicle flows and thus enables the management of vehicle passage at the proposed locations~\cite{b20}. In Marrakesh, Morocco \cite{b30}, 
a prototype capable of implementing a camera is shown in the windbreak of the vehicles, thus detecting if a driver slept while driving. In Tainan, Taiwan~\cite{b29}, LoRa used to  vehicle condition monitoring.

Another application class is urban monitoring. Authors in Yangon, Myanmar, and Tampines, Singapore \cite{b73}, have created sensors (with The Things Uno) that capture temperature, humidity, dust, and carbon dioxide in the air of cities. Air quality monitoring was also studied in La Plata, Argentina \cite{b83}. In the same scenario, some authors from Manouba, Tunisia \cite{b84}, focused on the feasibility of non-packet transmission with a LoRaWAN network implementing a sensor network in a city to monitor air quality. Monitoring the environment in cities was also the approach implemente in Hsinchu City, state of Taiwan~\cite{b86}. Urban pollutants control, in New York, USA~\cite{b31}, used LoRa to monitor the air in different neighborhoods.

In St. Petersburg, Russia~\cite{b23}, waste management applications were developed through LoRaWAN.  To assist in DHL cargo management at an airport in the Magdeburg region of Germany \cite{b74}, 
sensors were placed on the load conveyors (more than 1700 nodes) that communicate with a central gateway.

Authors in Brescia, Italy \cite{b71}, covered an area of 3.3 $km^{2}$ to evaluate variables in non-real time, such as the volume of dumps and monitoring of heaters. The maximum delay for both fixed and moving nodes was 250 ms due to the quality of the internet, which is sufficient for non-real time IoT applications. With similar hardware, in Selangor, Malaysia, a parking lot was implemented using LoRaWAN;  the project cost was considerably cheap considering its components~\cite{b21, b55}. In Kongens Lyngby, Denmark \cite{b66}, LoRa was used to monitor free boat spots in a harbor; the solution used ultrasonic sensors. In Brno, Czech Republic~\cite{b82}, a solution to assists in the management of large cars. Other studies, at Europe~\cite{b26, b28, b81} and Australia~\cite{b27}, have used general LoRa applications to characterize protocol behavior.

\item \textbf{Smart Grids}. Some solutions for Smart Grids were implemented in Brazil. In Santa Maria~\cite{b33}, 130 nodes were used to implement Smart Grids. 
Similar tests were conducted by~\cite{b72}.
In Campinas, Brazil~\cite{b34}, LoRaWAN and Mesh RF were compared for supply grid remote sensing.

In Meylan, France~\cite{b36}, it was showed the use of LoRaWAN with electric energy meters implementing nodes and gateways in Paris and analyzing the maximum reaches. In all 19 gateways separated by 1 km of each other were positioned and, in this scenario, 17 $km^{2}$ were covered using 11 B packets. The authors stressed, however, that since LoRa can be adapted for use in urban environments, the duty cycle issue is purely legal and not an impossibility in transmission, so it may or may not be respected. 
Another similar study~\cite{b87}, noted that if two packets are received simultaneously by a gateway, each coming from a node with the same SF and same frequency, the highest power packet will be decoded if it has 6 dB higher than the other packet.

In Sofia, Bulgaria \cite{b80}, low-cost hardware was used to obtain sensors information. In Ghent, Belgium~\cite{b35}, the temperature control in components related to power transmission lines could be done with the LoRa used an SX1276, component of commercial module RFM95. This service can be used to optimize energy deploy in smart cities~\cite{b59}. Similarly, applications with smart meters for water and gas consumption could be carried out by different authors in Tamilnadu and Uttar Pradesh, India \cite{b67, b68}.
In Hualien, Taiwan~\cite{b69}, authors were able to monitor the water quality in a lake using an ArduinoProMini sensor node for the gateway to communicate with a server via MQTT.

\item \textbf{Smart Farms}. Precision agriculture was the main application. In Campinas, Brazil~\cite{b37}, precise rural coverage information was studied under different gateway positioning heights, including using drones. Also in Torino, Italy \cite{b38}, and in Tronoh, Malaysia~\cite{b75, b76}. Authors in Kuala Lumpur and Tronoh, still in Malaysia \cite{b79}, have created an IoT network to monitor mushroom greenhouses. In Orissa, India~\cite{b52}, a circuit capable of monitoring solar panels via LoRaWAN.
In Skopje, Macedonia~\cite{b65}, a vineyard is monitored using sensor for air temperature, humidity, leaf water content and soil moisture.

Monitoring herds is another common application. In Montevideo, Uruguay \cite{b39}, each animal has a collar or ear tag with an accelerometer integrated with LoRa. In Ioannina, Grevena and Kavala, Greece \cite{b40}, a monitoring network for animals under LoRa transmission was developed by developing the nodes with Arduino SODAQ v2. An Android application (CowTrack) has also been developed to monitor animals. Monitor cows in fields was also studied in Baotou, China \cite{b64}.

Drones with LoRaWAN were used to identify the incidence of forest fires in Ruse, Bulgaria~\cite{b63}. In this scenario, 6000 sensor nodes were located at strategic points. Once detected by drones,  the occurrence of fire is notified for authorities.

\item \textbf{Health Care}. LoRa was used to monitor heart rate, respiration, blood fluid level, and other indicators, in Istanbul, Turkey~\cite{b61}, and Rome, Italy~\cite{b41}. In the studies in Sfax, Tunisia, and Marburg, Germany \cite{b42}, LoRa was used to monitor patients who are located far from health centers. 
In Belgrade, Serbia, and Ljubljana, Slovenia \cite{b43}, it was verified the implementation of remote sensing to monitor the health of athletes employing several sensors in the body.  In Karachi, Pakistan~\cite{b70}, authors have developed an embedded system that is capable of evaluating soldiers' health variables and sending data through LoRaWAN to a central data center.

\item \textbf{Location}. In Kongens Lyngby, Denmark \cite{b44}, LoRa was used to predict location with accuracy of 100 m for static cases. For moving cases, it is possible to report only approximations. The result was considered positive given that a network consisting of GPS and GSM consumes from 400 to 600 mA. This was also proven in other studies~\cite{b51,b85}.

\end{itemize}

LoRa and LoRaWAN have also been validated in applications for \textbf{University Campus} management (Lille, France~\cite{b45}, and Chiayi, Taiwan~\cite{b46}), \textbf{Industry} automation (Brescia, Italy, and Sundsvall, Sweden~\cite{b47}), and \textbf{Military} context (Moscow, Russia, and Almaty, Kazakhstan~\cite{b49} and in Bangkok, Thailand~\cite{b50}). In Maryland, USA \cite{b77}, authors used LoRaWAN communication during military tactical operations.

In all segments, system integration has used Message Queuing Telemetry Transport (MQTT)~\cite{b56}.
The work carried out in Plovdiv, Bulgaria~\cite{b48}, validated 
the operation of LoRa with MQTT.

\subsection{What are the main challenges that LoRa faces today?}

Based on the information obtained from the articles, the interference that the transmission via LoRa may cause in other signals is a point that deserves attention \cite{b49, b60}. It should be noted, however, that this finding was obtained at a specific frequency in the European scenario and does not necessarily represent interference in all other scenarios. In Brazil, for example, since LoRa operates at 915 MHz, the same types of situations shown by the authors are not relevant.

In continuous and real time data sensing (such as on ECG or athlete monitoring), LoRa technology may not be the right choice~\cite{b42, b43}. The authors' results, in this type of application, showed that the data rate for LoRa might not be sufficient.
It is also noticed that there are still few studies carried out in Brazil, justifying that the development of LoRa technology in the national territory lacks practical implementations.
Therefore, duty cycle and transfer rate are the performance limiters to adhere to LoRa as the best technology for remote sensing and IoT. The integration of LoRa with cellular networks would bring numerous benefits \cite{b62}.


\section{Security} \label{sec}

Most of LoRaWAN applications did not analyzed cybersecurity risks. In this particular case, LoRaWAN, from version 1.0 (2015) to version 1.1 (2017), has made significant security improvements. An important point is that LoRaWAN protocol does not support firmware update, so that components of version 1.0 and 1.1 may compose a application~\cite{03_security_LoRaWANv1.1_scenarios}. This compatibility scenario shall be considered in cybersecurity analysis.

Cyberattacks can cause significant business impact to organizations that do not have the suitable cybersecurity mechanisms in place. Successfully attacks can affect the organization reputation, may result in sentitive data loss or exposure of intellectual properties.

Formal analysis of LoRaWAN specification version 1.1 showed that --- if all the requirements are followed --- there will be no cybersecurity vulnerabilities~\cite{01_formal_security_analysis_LoRaWAN}; it does not consider the inherent vulnerabilities of wireless communication. In that case the cybersecurity vulnerabilities can occurs mainly in design and deploy process. Most of the vulnerabilities of LoRaWAN are related to OTAA join procedure (device activation process)~\cite{92_DoS_lorawan, 109_key_management_healthcare}.

\section{Conclusion} \label{concl}
This work observed that the technology was satisfactorily applied for most of the projects studied. The applications were  dependent on the experimental environment, being influenced by topology, SF used, urban density, hardware (dedicated or prototype printed circuit), antenna, mode package management, and other factors. Such details were not always adequately informed in the articles.

LoRa demonstrated  several positive points (mainly, greater distance range and low energy consumption), making it a very viable option for IoT in many scenarios.  As future works, we intend to complement the bibliographic review by examining the categories of theoretical and experimental articles. Besides, it is desired to deepen the comparisons on results of latency and energy consumption. Finally, we are currently working on a low-cost LoRa solution for monitoring weather events; in near future, we hope to provide another practical application of LoRa.

\bibliographystyle{sbc}
\bibliography{sbc-template}

\end{document}